\documentclass[aps, preprint, showkeys, nofootinbib]{revtex4}
\usepackage{amsmath}
\usepackage{amsfonts}
\usepackage{amssymb}
\usepackage{graphicx}%
\providecommand{\U}[1]{\protect\rule{.1in}{.1in}}
\begin{document}

\preprint{ }
\title{Homogeneous Cosmological Models in Weyl's Geometrical Scalar--Tensor Theory}
\author{A. Barros}
\affiliation{Centro de Desenvolvimento Sustent\'{a}vel do Semi\'{a}rido, Universidade
Federal de Campina Grande, 58540-000, Sum\'{e}, PB, Brazil.}
\affiliation{atbarros@ufcg.edu.br}
\author{C. Romero}
\affiliation{Departamento de F\'{\i}sica, Universidade Federal da Para\'{\i}ba. C.P. 5008,
58059-970, Jo\~{a}o Pessoa, PB, Brazil.}
\affiliation{cromero@fisica.ufpb.br}

\begin{abstract}
In this paper, we consider homogeneous cosmological solutions in the context
of the Weyl geometrical scalar--tensor theory. Firstly, we exhibit an
anisotropic Kasner type solution taking advantage of some similarities between
this theory and the Brans--Dicke theory. Next, we consider an isotropic model
with a flat spatial section sourced by matter configurations described by a
perfect fluid. In this model, we obtain an analytical solution for the stiff
matter case. For other cases, we carry out a complete qualitative analysis
theory to investigate the general behaviour of the solutions, presenting some
possible scenarios. In this work, we do not consider the presence of the
cosmological constant nor do we take any potential of the scalar field into
account. Because of this, we do not find any solution describing the
acceleration of the universe.

\end{abstract}
\keywords{Weyl geometry; Scalar-tensor theory; Cosmological models.}\maketitle

\section{Introduction}

As is well known, scalar--tensor theories of gravity were proposed some years
ago by Jordan~\cite{jor59}, and~Brans and Dicke~\cite{Bra61}. Later, they were
extended in a more general framework~\cite{ber68,wag70,nor70}. In~fact, they
represent a generalization of the simplest scalar--tensor theory of gravity
which is the Brans--Dicke theory~\cite{far04,wil14}. In~general scalar--tensor
theories of gravity, the~gravitational field is not described only by the
usual tensor field $g_{\mu\nu}$ of general relativity. In~addition to this, we
have one or several long-range scalar fields which also mediate gravitational interaction.

Scalar--tensor theories of gravity have been a subject of renewed interest.
Certainly, one motivation for this is the belief that, at~least at
sufficiently high energy scales, gravity becomes scalar--tensorial in
nature~\cite{dam94} and, therefore, these theories are important in the very
early Universe. On~the other hand, two important theoretical developments have
been achieved such as, for~example, unification models based on superstrings,
which naturally associate long-range scalar partners to the usual tensor
gravity of Einstein theory~\cite{gre87}. Another motivation for the
investigation of scalar--tensor theories is that inflationary cosmology in
this framework seems to solve the fine-tuning problem and, in this way, give
us a mechanism of terminating inflationary eras~\cite{la89}. Apart from the
solution of this problem, the~scalar--tensor theories by themselves have
direct implications for cosmology and for experimental tests of the
gravitational interaction~\cite{dam00} and for this reason, they are relevant
in the investigation of the early~Universe.

Among alternative theories of gravity, scalar--tensor theories are perhaps the
most popular ones. As~we have pointed out before, in~these theories,
gravitational effects are described by both a metric field $g_{\mu\nu}$ and a
scalar field $\Phi$. A~well-known example is the Brans--Dicke
theory~\cite{Bra61,tah21}, in~which the geometry of the underlying space-time
manifold is assumed to be Riemannian, and~the scalar field replaces the
gravitational constant being interpreted as the inverse of a varying
gravitational coupling parameter. In~addition to the reasons mentioned above,
the~scalar--tensor theories are studied because they admit key ingredients of
string theories, such as a dilaton-like gravitational scalar field that has a
non-minimal coupling to the curvature~\cite{mor11}. On~the other hand,
a~different approach, in~which the scalar field appears as part of the
space-time geometry, namely, the~Weyl geometrical scalar--tensor theory, has
been discussed recently in the literature~\cite{rom14}. Indeed, in~this new
approach, one considers the space-time structure as a very special case of the
framework adopted in the original Weyl unified field theory~\cite{wey18,wey52}%
, the~geometrical space-time structure being that of a \textit{Weyl integrable space-time} (WIST) \cite{rom08,rom09,rom10,rom11,rom12}. It is important to remark that
other gravity theories in which a scalar field plays a geometrical role have
also been proposed~\cite{Fonseca,Fonseca1,Fonseca2}.

Recently, some theoretical aspects concerning the Weyl geometrical
scalar--tensor theory have been studied, in~particular the behaviour of the
solutions when $\omega$, the~scalar field's coupling constant, goes to
infinity~\cite{barros}. The~investigation of cosmological vacuum models for
different scalar potentials has also been carried out~\cite{rom16}. In~the
present article, we extend this research to include anisotropic models of
Kasner type. Here, we take advantage of some similarities between vacuum
solutions of the Weyl geometrical scalar--tensor theory and those coming from
the Brans--Dicke theory. We also examine cosmological solutions in the
presence of matter, a~scenario that has not yet been investigated in Weyl
geometrical scalar--tensor theory, and~at the same time, we compare the
results obtained with similar solutions already known from general relativity
and the Brans--Dicke~theory.

The paper is organized as follows. In~Section~II, we briefly
review Weyl's original theory, which inspired the geometrical scalar--tensor
approach. In~Section~III, the~field equations of the Weyl
geometrical scalar--tensor theory are obtained. Then, a~Kasner type solution
is exhibited in Section~IV, while in
Section~V we work with a homogeneous and
isotropic cosmological model having a perfect fluid as a source, such that we
find an analytical solution for the stiff matter case and we study the other
cases using the qualitative analysis of dynamical systems. Finally,
Section~VI is devoted to our~conclusions.

\section{Weyl's~Theory}

In the first scalar--tensor theories, the~so-called Jordan--Brans--Dicke
theories, it is assumed, as~in general relativity, that the space-time
geometry is purely Riemannian. On~the other hand, if~the Palatini variational
method is applied to deduce the field equations from the action, then in a
large class of scalar--tensor theories, a~non-Riemannian compatibility
condition between the metric and the affine connection appears naturally (for
a more general result, see~\cite{bur98}). In~this way, we have a theory that
establishes the space-time geometry from first principles, that is,
the~space-time manifold is dynamically generated by the choice of the
particular coupling of the scalar field in the gravitational sector. In~the
case where the action is that of the Brans--Dicke theory, this procedure leads
to the so-called \textit{Weyl integrable space-times}, a~particular version of the geometry conceived by H. Weyl in his attempt to
unify gravity and electromagnetism~\cite{wey18}. Note, however, that here, it
is the scalar field that is being~geometrized.

It is true that the Weyl geometry is one of the simplest generalizations of
Riemannian geometry, in~which the Riemannian compatibility condition between
the metric and the affine connection is weakened. This was an ingenious way
that Weyl devised to introduce a covariant vector field $\sigma_{\mu}$ in the
geometry, which bears a great similarity with the electromagnetic
four-potential. Weyl went on and introduced the second-order tensor $F_{\mu
\nu}=\partial_{\mu}\sigma_{\nu}-\partial_{\nu}\sigma_{\mu}$, which he
interpreted as representing another kind of curvature, namely,
the~\textit{length curvature}. As~a consequence of this modification in the
Riemannian compatibility condition, the~covariant derivative of the metric
tensor does not vanish, as~in Riemannian geometry, and~the length of vectors
when parallel transported along a curve may change. However, such theory
suffered from a severe criticism by Einstein, who objected that the
nonintegrability of length implies that the rate at which a clock measures
time, i.e.,~its clock rate, in~this case would depend on the past history of
the clock. As~a consequence of this fact, spectral lines with sharp
frequencies would not appear and the spectrum of neighbouring elements of the
same kind would be different~\cite{goe04}. This became known in the literature
as the second clock effect (incidentally, the~first clock effect refers to the
well-known effect corresponding to the ``twin paradox'', which is predicted by
special and general relativity theories).

Weyl's new compatibility condition is given by $\nabla_{\alpha}g_{\mu\nu
}=\sigma_{\alpha}g_{\mu\nu}$, and~is easily verified that this condition is
invariant under the conformal transformation $g_{\mu\nu}\rightarrow\bar
{g}_{\mu\nu}=e^{f}g_{\mu\nu}$ carried out simultaneously with the
\textit{gauge }transformation $\sigma_{\mu}\rightarrow\bar{\sigma}_{\mu
}=\sigma_{\mu}+\partial_{\mu}f$, where $f$ is an arbitrary scalar function.
The~discovery of this new symmetry is now considered by some authors as the
birth of\ modern gauge theories~\cite{ale13}. Now, if~$F_{\mu\nu}=0$ (null
second curvature), which is equivalent to say that the one-form $\sigma$ is
closed ($d\sigma=0$), then there is no electromagnetic field. In~this case, we
know that, from~Poincar\'{e}'s lemma~\cite{tu11}, it follows that there exists
a scalar field $\phi$, such that $\sigma_{\mu}=\partial_{\mu}\phi$, and,
instead of a vector field $\sigma$, we are left with a scalar field $\phi$,
which, in~addition to the metric, is the fundamental object that characterizes
the geometry. A~space-time endowed with this particular version of Weyl's
geometry came to be known as a\textit{Weyl integrable space-time}.

\section{The Field~Equations}

As we have already mentioned, in~the Weyl geometrical scalar--tensor theory,
the underlying space-time manifold is that of a Weyl integrable
space-time~\cite{rom08}. In~this sense, the~Weyl nonmetricity condition
involves a purely geometrical scalar field $\phi$ and is explicitly given
by~\cite{rom14}%
\begin{equation}
\triangledown_{\alpha}g_{\mu\nu}=g_{\mu\nu}\phi_{,\alpha}. \label{1}%
\end{equation}

Moreover, one can define the Weyl connection, whose coefficients in a local
coordinate basis read
\begin{equation}
\Gamma_{\mu\nu}^{\alpha}=\{_{\mu\nu}^{\alpha}\}-\frac{1}{2}g^{\alpha\beta
}(g_{\beta\mu}\phi_{,\nu}+g_{\beta\nu}\phi_{,\mu}-g_{\mu\nu}\phi_{,\beta}),
\label{2}%
\end{equation}
with $\{_{\mu\nu}^{\alpha}\}$ representing the usual Christoffel~symbols.

In turn, the~field equations of the Weyl geometrical scalar--tensor theory can
be written as~\cite{barros}%
\begin{align}
G_{\mu\nu}  &  =-\frac{(\omega-\frac{3}{2})\ }{\Phi^{2}\ }\left(  \Phi_{,\mu
}\Phi_{,\nu}-\frac{g_{\mu\nu}}{2}\Phi_{,\alpha}\Phi^{,\alpha}\right)
\nonumber\\
&  -\frac{1}{\Phi}(\Phi_{,\mu;\nu}-g_{\mu\nu}\square\Phi\ )-\frac{g_{\mu\nu}%
}{2\Phi}V(\Phi)-8\pi T_{\mu\nu}, \label{3}%
\end{align}%
\begin{equation}
\square\Phi\ =\frac{1}{\omega}\left(  -\frac{1}{2}\frac{dV}{d\Phi}\Phi
+V(\Phi)\right)  , \label{4}%
\end{equation}
where here, we are using the field variable $\Phi=e^{-\phi}$, $\omega=const$,
$V(\phi)$ corresponds to the scalar field potential, and~$T_{\mu\nu}$
represents the Weyl invariant energy--momentum tensor of the matter fields. We
denote by $G_{\mu\nu}$ and $\square$ the Einstein tensor and the d'Alembertian
operator, respectively, defined with respect to the Christoffel symbols.
If~$V(\Phi)=2\Lambda\Phi$, one can introduce the cosmological constant
$\Lambda$. However, let us take $\Lambda=0$, and~then the field equations are
given by%
\begin{align}
G_{\mu\nu}  &  =-\frac{W\ }{\Phi^{2}\ }\left(  \Phi_{,\mu}\Phi_{,\nu}-\frac
{1}{2}g_{\mu\nu}\Phi_{,\alpha}\Phi^{,\alpha}\right) \nonumber\\
&  -\frac{1}{\Phi}\Phi_{,\mu;\nu}-8\pi T_{\mu\nu}, \label{5}%
\end{align}%
\begin{equation}
\square\Phi\ =0, \label{6}%
\end{equation}
where $W=\omega-\frac{3}{2}$. Additionally, we can obtain from (\ref{5}) and
(\ref{6}) that
\begin{equation}
R_{\mu\nu}=-8\pi T_{\mu\nu}+\frac{8\pi T}{2}g_{\mu\nu}-\frac{W}{\Phi^{2}}%
\Phi_{,\mu}\Phi_{,\nu}-\frac{\Phi_{,\mu;\nu}}{\Phi}, \label{7}%
\end{equation}
with $R_{\mu\nu}$ denoting the Ricci tensor and $T=g_{\mu\nu}T^{\mu\nu}$.
Equations~(\ref{6}) and (\ref{7}) constitute the field equations we use in the following.

\section{Kasner Type~Solution}

As is well known, the~Kasner metric was obtained by the mathematician E.
Kasner in 1921 and~represents an exact solution to Einstein's field equations.
It describes an anisotropic universe without matter, that is, it is a vacuum
solution. Historically, interest in the Kasner solution came from the fact
that, although~it may have a singularity (``big bang'' or a ``big crunch''),
an isotropic expansion or contraction of space is not allowed, and~this led to
the generic singularity studies, the~so-called BKL singularities~\cite{tip79}.

The Kasner type solution in the Brans--Dicke theory of gravity is given
by~\cite{Bra61,rub72}
\begin{equation}
ds^{2}=dt^{2}+R_{1}^{2}dx^{2}+R_{2}^{2}dy^{2}+R_{3}^{2}dz^{2}, \label{8}%
\end{equation}
with%
\begin{equation}
R_{i}=r_{i}(at+b)^{\frac{p_{i}}{1+C}},\text{ } \label{9}%
\end{equation}
($i=1,2,3$) and the Brans--Dicke scalar field%
\begin{equation}
\varphi=\varphi_{0}(at+b)^{\frac{C}{1+C}}, \label{10}%
\end{equation}
where $a$, $b$, $r_{i}$, and $\varphi_{0}$ are constants. The~relations $%
%TCIMACRO{\tsum }%
%BeginExpansion
{\textstyle\sum}
%EndExpansion
p_{i}=1$ and%
\begin{equation}%
%TCIMACRO{\tsum }%
%BeginExpansion
{\textstyle\sum}
%EndExpansion
p_{i}^{2}=1-C(\omega C-2) \label{11}%
\end{equation}
between the constants $p_{i},C$ and the scalar field coupling constant
$\omega$ are also~satisfied.

The space-time given by (\ref{8}) corresponds to a homogeneous universe,
without matter and rotation, with~distinct expansions along the three
orthogonal axes, which reflects anisotropy. Note that if $a=1$ and $b=0$,
Equations~(\ref{9}) and (\ref{10}) may be written as%
\begin{equation}
R_{i}=r_{i}t^{\frac{p_{i}}{1+C}},\text{ } \label{12}%
\end{equation}%
\begin{equation}
\varphi=\varphi_{0}t^{\frac{C}{1+C}}.\text{ } \label{13}%
\end{equation}

In order to obtain a solution in the Weyl geometrical scalar--tensor theory,
let us consider the following result: a vacuum solution of the Weyl
geometrical scalar--tensor theory can be found if we make the change
$\omega\rightarrow W=\omega-3/2$ in the correspondent vacuum solution of the
Brans--Dicke theory. In~fact, the~two theories are not physically equivalent
given that in Weyl's geometrical scalar--tensor theory test particles follow
affine Weyl geodesics (autoparallels) and not metric geodesics as in the case
of the Brans--Dicke theory. Nonetheless, there is a formal equality between
the vacuum field equations of the two theories~\cite{rom14}.

Thus, the~Kasner type solution in the Weyl geometrical scalar--tensor theory
is given by Equation~(\ref{12}) and%
\begin{equation}
\Phi=\Phi_{0}t^{\frac{C}{1+C}},\text{ } \label{14}%
\end{equation}
where $%
%TCIMACRO{\tsum }%
%BeginExpansion
{\textstyle\sum}
%EndExpansion
p_{i}=1$ and%
\begin{equation}%
%TCIMACRO{\tsum }%
%BeginExpansion
{\textstyle\sum}
%EndExpansion
p_{i}^{2}=1-C(WC-2)=1-C\left[  \left(  \omega-\frac{3}{2}\right)  C-2\right]
. \label{15}%
\end{equation}

Now, if~we choose $C=\dfrac{2}{W}$, it follows that%
\begin{equation}%
%TCIMACRO{\tsum }%
%BeginExpansion
{\textstyle\sum}
%EndExpansion
p_{i}^{2}=1. \label{16}%
\end{equation}

Furthermore, (\ref{12}) and (\ref{14}) become%
\begin{equation}
R_{i}=r_{i}t^{\frac{Wp_{i}}{W+2}}=r_{i}t^{\left[  \left(  \omega-3/2\right)
/\left(  \omega+1/2\right)  \right]  p_{i}}, \label{17}%
\end{equation}%
\begin{equation}
\Phi=\Phi_{0}t^{\frac{2}{W+2}}=\Phi_{0}t^{\left[  2/\left(  \omega+1/2\right)
\right]  }. \label{18}%
\end{equation}

In the limit $\omega\rightarrow\infty$, (\ref{17}) and (\ref{18}) tend to%
\begin{equation}
R_{i}=t^{p_{i}},\text{ } \label{19}%
\end{equation}%
\begin{equation}
\Phi=\Phi_{0},\text{ } \label{20}%
\end{equation}
where we have taken $r_{i}=1$. On~the other hand, from~(\ref{1}) and (\ref{2})
we find that
\begin{equation}
\nabla_{\alpha}g_{\mu\nu}=-g_{\mu\nu}\left(  \frac{\Phi_{,\alpha}}{\Phi
}\right)  , \label{21}%
\end{equation}%
\begin{equation}
\Gamma_{\mu\nu}^{\alpha}=%
%TCIMACRO{\QTATOPD{\{}{\}}{\alpha}{\mu\nu}}%
%BeginExpansion
\genfrac{\{}{\}}{0pt}{1}{\alpha}{\mu\nu}%
%EndExpansion
+\frac{1}{2\Phi}g^{\alpha\beta}\left(  g_{\beta\mu}\Phi_{,\nu}+g_{\beta\nu
}\Phi_{,\mu}-g_{\mu\nu}\Phi_{,\beta}\right)  , \label{22}%
\end{equation}
by considering the scalar field in the form $\Phi=e^{-\phi}$. Thus, when
$\omega\rightarrow\infty$, the~space-time geometry becomes Riemannian as we
have
\begin{equation}
\nabla_{\alpha}g_{\mu\nu}=0,\text{ \ and\ \ }\Gamma_{\mu\nu}^{\alpha}=%
%TCIMACRO{\QTATOPD{\{}{\}}{\alpha}{\mu\nu}}%
%BeginExpansion
\genfrac{\{}{\}}{0pt}{1}{\alpha}{\mu\nu}%
%EndExpansion
. \label{23}%
\end{equation}

Therefore, also taking into account (\ref{19}) and (\ref{20}), the~Kasner
solution of general relativity is recovered in this~limit.

\section{A Perfect Fluid Cosmological~Model}

The Friedmann--Robertson--Walker metric with a flat spatial section is given
by
\begin{equation}
ds^{2}=dt^{2}-R^{2}(t)\left[  dr^{2}+r^{2}\left(  d\vartheta^{2}+\sin
^{2}\vartheta d\chi^{2}\right)  \right]  , \label{24}%
\end{equation}
where $R(t)$ denotes the scale factor. In~this cosmological model, the~matter
content is a perfect fluid represented by the energy--momentum tensor%
\begin{equation}
T_{\mu\nu}=\left(  p+\rho\right)  u_{\mu}u_{\nu}-pg_{\mu\nu}, \label{25}%
\end{equation}
with $p=\lambda\rho$, $0\leq\lambda\leq1$, $p$ being the thermodynamic
pressure, $\rho$ the energy density, and $u_{\mu}=\left(  1,0,0,0\right)  $
the four-velocity vector field. Then, field Equations~(\ref{6}) and (\ref{7})
reduce to%
\begin{equation}
\frac{3\ddot{R}}{R}=-4\pi\rho\left(  1+3\lambda\right)  -W\frac{\dot{\Phi}%
^{2}}{\Phi^{2}}-\frac{\ddot{\Phi}}{\Phi}, \label{26}%
\end{equation}%
\begin{equation}
\frac{\ddot{R}}{R}+\frac{2\dot{R}^{2}}{R}=4\pi\rho\left(  1-\lambda\right)
-\frac{\dot{R}\dot{\Phi}}{R\Phi}, \label{27}%
\end{equation}%
\begin{equation}
\frac{\ddot{\Phi}}{\Phi}+\frac{3\dot{R}\dot{\Phi}}{R\Phi}=0. \label{28}%
\end{equation}

The dot means differentiation with respect to time. Moreover, due to the
assumption of spatial homogeneity, the~scalar field $\Phi$ is supposed to be a
function of $t$ only. Additionally, with~the definitions $\theta=\frac
{3\dot{R}}{R}$ and $\Psi=\frac{\dot{\Phi}}{\Phi}$, one can express
(\ref{26})--(\ref{28}) in the form%
\begin{equation}
\dot{\theta}=-\frac{\theta^{2}}{3}-4\pi\rho\left(  1+3\lambda\right)  -\left(
W+1\right)  \Psi^{2}-\dot{\Psi}, \label{29}%
\end{equation}%
\begin{equation}
\dot{\theta}=-\theta^{2}+12\pi\rho\left(  1-\lambda\right)  -\theta\Psi,
\label{30}%
\end{equation}%
\begin{equation}
\dot{\Psi}=-\Psi^{2}-\theta\Psi. \label{31}%
\end{equation}

By combining (\ref{29})--(\ref{31}), we can derive the equation%
\begin{equation}
\frac{\theta^{2}}{3}-\frac{W\Psi^{2}}{2}+\theta\Psi=8\pi\rho. \label{32}%
\end{equation}

After some calculations and by using Equations~(\ref{5}) and (\ref{6}), it is
easy to show that%
\begin{equation}
T_{\text{ \ \ };\nu}^{\mu\nu}=\frac{T}{2}\frac{\Phi^{,\mu}}{\Phi}-\frac
{\Phi_{,\nu}}{\Phi}T_{\text{ \ \ }}^{\mu\nu}, \label{33}%
\end{equation}
which reduces to%
\begin{equation}
\dot{\rho}=-\left[  \left(  1+\lambda\right)  \theta+\left(  \frac{1+3\lambda
}{2}\right)  \Psi\right]  \rho\label{34}%
\end{equation}
in the context of the cosmological model~considered.

\subsection{Stiff Matter~Solution}

Next, we obtain the equations of a dynamic system which lead us to carry out a
rich analysis of the solutions. For~this purpose, let us consider the
following equation, which results from (\ref{29})--(\ref{31}):%
\begin{equation}
\dot{\theta}=-\frac{\left(  1+\lambda\right)  }{2}\theta^{2}+\frac{\left(
1-3\lambda\right)  }{2}\theta\Psi-\frac{3W\left(  1-\lambda\right)  }{4}%
\Psi^{2}. \label{35}%
\end{equation}

This equation, together with (\ref{31}), constitutes a homogeneous autonomous
planar dynamic system. It is important to note that the solutions of this
system, $\theta(t)$ and $\Psi(t)$, must necessarily satisfy the constraint
imposed by Equation~(\ref{32}).

Cosmological scenarios modelled by stiff matter have been investigated
recently, particularly in connection with the problem of dark
matter~\cite{pie15}. Now, let us consider the stiff matter case in the
geometrical scalar--tensor theory. Then, it follows from (\ref{35}), in~the
case known as stiff matter ($\lambda=1$), that%
\begin{equation}
\dot{\theta}=-\theta^{2}-\theta\Psi. \label{36}%
\end{equation}

Clearly, an~immediate solution of the system of Equations~(\ref{31}) and
(\ref{36}) is given by $\Psi=-\theta$, which leads to the particular solution
$(\theta=\theta_{0}$ $,$ $\Psi=\Psi_{0})$, $\theta_{0}$ and $\Psi_{0}$ being
constants. Hence, we have $\frac{3\dot{R}}{R}=\theta_{0}$, $\frac{\dot{\Phi}%
}{\Phi}=\Psi_{0}$,\ which then leads to%
\begin{equation}
R(t)=R_{0}\exp\left(  \frac{\theta_{0}t}{3}\right)  , \label{38}%
\end{equation}%
\begin{equation}
\Phi(t)=\Phi_{0}\exp\left(  \Psi_{0}t\right)  , \label{39}%
\end{equation}
where $R_{0}$ and $\Phi_{0}$ are constants, which we recognize as a de Sitter
type solution, with~the scalar field also having an exponential behaviour.
Furthermore, from~(\ref{32}), we can find%
\begin{equation}
\rho=-\frac{\left(  3W+4\right)  }{48\pi}\theta_{0}^{2}. \label{40}%
\end{equation}

Now, by~defining $\alpha=\theta+\Psi\neq0$, let us find the general solution
for the stiff matter case. To~do this, one can add Equations~(\ref{31}) and
(\ref{36}) to~obtain%
\begin{equation}
\dot{\alpha}+\alpha^{2}=0, \label{41}%
\end{equation}
whose solution is%
\begin{equation}
\alpha=\frac{1}{t+D}, \label{42}%
\end{equation}
where $D$ is a constant. In~turn, from~Equation~(\ref{34}) with $\lambda=1$,
it follows that%
\begin{equation}
\dot{\rho}+2\alpha\rho=0, \label{43}%
\end{equation}
whose solution is%
\begin{equation}
\rho=\frac{\rho_{0}}{\left(  t+D\right)  ^{2}}, \label{44}%
\end{equation}
with $\rho_{0}$ constant.

To obtain the expression of $\theta$, let us consider (\ref{32}) and
(\ref{44}) and use that \mbox{$\Psi=\alpha-\theta=\frac{1}{t+D}-\theta$}.
In~this way we are led to
\begin{equation}
\theta=\frac{B}{t+D}, \label{45}%
\end{equation}%
\begin{equation}
\Psi=\frac{1-B}{t+D}, \label{46}%
\end{equation}
where%
\begin{equation}
B=\frac{3(W+1)\pm\sqrt{3\left[  \left(  3+2W\right)  -16\pi\rho_{0}\left(
4+3W\right)  \right]  }}{4+3W}, \label{47}%
\end{equation}
while the condition $\left(  3+2W\right)  -16\pi\rho_{0}\left(  4+3W\right)
\geq0$ is required to be~satisfied.

On the other hand, if~we replace (\ref{45}) and (\ref{46}) in (\ref{32}), we
obtain%
\begin{equation}
\rho=\frac{1}{8\pi\left(  t+D\right)  ^{2}}\left[  \frac{B^{2}}{3}-\frac{W}%
{2}\left(  1-B\right)  ^{2}+B\left(  1-B\right)  \right]  . \label{49}%
\end{equation}

The solutions for the scale factor and the scalar field can be obtained by
integrating the expressions $\theta=\frac{3\dot{R}}{R}$ and $\Psi=\frac
{\dot{\Phi}}{\Phi}$, giving the following:%
\begin{equation}
R(t)=R_{0}\left(  t+D\right)  ^{B/3}, \label{50}%
\end{equation}%
\begin{equation}
\Phi(t)=\Phi_{0}\left(  t+D\right)  ^{1-B}, \label{51}%
\end{equation}
with $R_{0}$ and $\Phi_{0}$ being~constants.

It should be noted that the constant $B$ can also be written as%
\begin{equation}
B=1-\frac{1}{4+3W}\pm\frac{\sqrt{3\left[  \left(  3+2W\right)  -16\pi\rho
_{0}\left(  4+3W\right)  \right]  }}{4+3W}. \label{52}%
\end{equation}

Thus, for~a large $W$, we obtain%
\begin{equation}
B=1-\frac{1}{3W}\pm\sqrt{\frac{1}{3W}\left(  2-48\pi\rho_{0}\right)  }\text{
}. \label{53}%
\end{equation}

Let us now consider that $\rho_{0}=\frac{f(W)}{24\pi}$, $f(W)$ being a
function which tends to one when $W$ is large. Therefore, $B$ takes the form%
\begin{equation}
B=1-\frac{1}{3W}=1+O\left(  \frac{1}{W}\right)  . \label{54}%
\end{equation}

Under these conditions, one can obtain from (\ref{51}) that%
\begin{equation}
\Phi(t)=\Phi_{0}+O\left(  \frac{1}{W}\right)  . \label{55}%
\end{equation}

When $\Phi$ behaves as in (\ref{55}) for a large $W$, it has been verified
that any vacuum solution of the Weyl geometrical scalar--tensor theory reduces
to the corresponding general relativistic solution in the limit $W\rightarrow
\infty$ \cite{barros}. This fact also occurs here, since Equations~(\ref{50}%
)~and~(\ref{51}) become equal to the Einstein solution%
\begin{equation}
R(t)=R_{0}t^{1/3}, \label{56}%
\end{equation}%
\begin{equation}
\Phi=\Phi_{0}, \label{57}%
\end{equation}
for $W\rightarrow\infty$ (we take $D=0$). Naturally, the~geometry of the
space-time becomes Riemannian, according to Equations~(\ref{21})--(\ref{23}).

\subsection{Qualitative Analysis for $\lambda\neq1$}

For values of the parameter $\lambda$ in the interval $0\leq\lambda<1$, we use
the qualitative analysis theory~\cite{andronov}, by~which many of the general
characteristics of the integral solutions of the system can be studied without
working out explicit solutions $\theta(t)$ and $\Psi(t)$. For~this purpose,
let us start by writing Equations~(\ref{31}) and (\ref{35}) of the dynamic
system as%
\begin{equation}
\dot{\theta}=F(\theta,\Psi)=-\frac{\left(  1+\lambda\right)  }{2}\theta
^{2}+\frac{\left(  1-3\lambda\right)  }{2}\theta\Psi-\frac{3W\left(
1-\lambda\right)  }{4}\Psi^{2}, \label{58}%
\end{equation}%
\begin{equation}
\dot{\Psi}=H(\theta,\Psi)=-\Psi^{2}-\theta\Psi. \label{59}%
\end{equation}

An equilibrium point of the system, i.e.,~a solution that occurs when
$F(\theta,\Psi)=H(\theta,\Psi)=0$, is the origin of the phase plane, the~point
$M$ $(\theta=0,\Psi=0)$. This solution represents Minkowski's space-time,
being the only finite equilibrium point that is significant in the~system.

In the qualitative analysis of solutions of Equations~(\ref{58}) and
(\ref{59}), one must construct the phase diagrams. For~this, we make use of
the Poincar\'{e} compactification method, which projects the phase plane into
a sphere. A~second mapping, in~turn, projects this sphere orthogonally onto a
disk, whose circumference represents the infinity of the initial phase
plane~\cite{andronov}.

\subsection{Invariant Rays and Regions of Negative Energy~Density}

Initially, in~our analysis, we obtain the invariant rays of the dynamic system
defined above. For~this, let us make the change of variables $\theta
=r\cos\beta$ and $\Psi=r\sin\beta$, $r$ and $\beta$ being polar coordinates of
the plane. In~this way, we find%
\begin{equation}
\dot{\theta}=r^{2}\left[  -\frac{\left(  1+\lambda\right)  }{2}\cos^{2}%
\beta+\frac{\left(  1-3\lambda\right)  }{2}\cos\beta\sin\beta-\frac{3W\left(
1-\lambda\right)  }{4}\sin^{2}\beta\right]  =r^{2}\overline{F}(\beta),
\label{60}%
\end{equation}%
\begin{equation}
\dot{\Psi}=r^{2}\left[  -\sin^{2}\beta-\cos\beta\sin\beta\right]
=r^{2}\overline{H}(\beta). \label{61}%
\end{equation}

Now, from~the relations between the variables $\theta$, $\Psi$, $r$, $\beta$,
and Equations~(\ref{60})~and~(\ref{61}), it can be shown that%
\begin{equation}
\dot{\beta}=r\left(  -\overline{F}(\beta)\sin\beta+\overline{H}(\beta
)\cos\beta\right)  . \label{62}%
\end{equation}

Next, we obtain the invariant rays, which, by~definition, consist of solutions
where the ratio $\frac{\Psi}{\theta}=\tan\beta=const$. Thus, putting
$\dot{\beta}=0$ in expression (\ref{62}) leads to%
\begin{equation}
\tan\beta=\frac{\overline{H}(\beta)}{\overline{F}(\beta)}. \label{63}%
\end{equation}

Again, with~the help of Equations~(\ref{60}) and (\ref{61}), it follows from
(\ref{63}) that%
\begin{equation}
\tan\beta\left(  \frac{W}{2}\tan^{2}\beta-\tan\beta-\frac{1}{3}\right)  =0.
\label{64}%
\end{equation}

\begin{figure}[h]
\begin{center}
\includegraphics[width=7.5 cm]{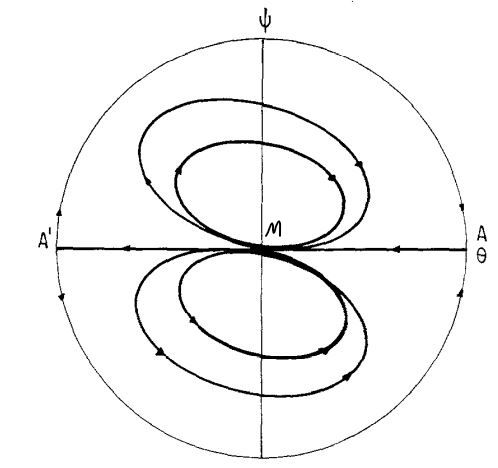}
\end{center}
\unskip
\caption{\centering{$\ \ W<-\frac{3}{2}$ ($\omega<0$).}\label{fig1}}
\end{figure}

\begin{figure}[h]
\begin{center}
\includegraphics[width=7.5 cm]{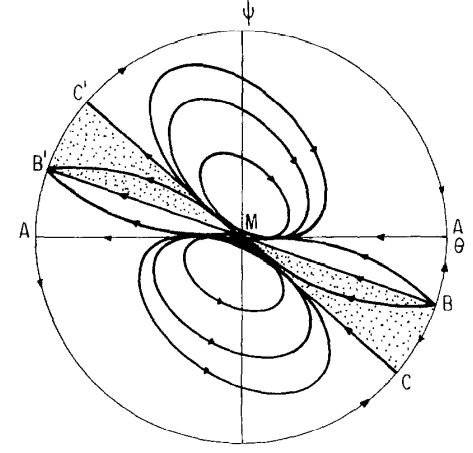}
\end{center}
\unskip
\caption{\centering{$\ \ -\frac{3}{2}<W<-\frac{4}{3}$ ($0<\omega<\frac{1}{6}$).}\label{fig2}}
\end{figure}

For $W<-\frac{3}{2}$, the~roots of (\ref{64}) are $\beta_{1}=0$ and $\beta
_{2}=\pi$. The~solutions representing these invariant rays appear in phase
diagrams such as curves $AM$ and $MA^{\prime}$, respectively (see
Figure~\ref{fig1}, for example). When $W>-\frac{3}{2}$, in~addition to the
roots $\beta_{1}$ and $\beta_{2}$ already mentioned, there are four more:%
\begin{equation}
\beta_{3}=\tan^{-1}\left[  -\frac{3}{2}\left(  1+\sqrt{1+\frac{2W}{3}}\right)
\right]  ^{-1},\text{ \ \ \ }\beta_{4}=\beta_{3}+\pi, \label{65}%
\end{equation}%
\begin{equation}
\beta_{5}=\tan^{-1}\left[  -\frac{3}{2}\left(  1-\sqrt{1+\frac{2W}{3}}\right)
\right]  ^{-1},\text{ \ \ \ }\beta_{6}=\beta_{5}+\pi, \label{66}%
\end{equation}
which correspond to the curves $BM$, $MB^{\prime}$, $CM$, and $MC^{\prime}$,
respectively (see Figure~\ref{fig2}, for~instance). These invariant rays
depend on $W$ and, as~its value increases, the~following behaviour is
observed: the line $BB^{\prime}$ rotates anticlockwise approaching the
$\theta$-axis, while the line $CC^{\prime}$ moves clockwise tending to make an
angle of $-180%
%TCIMACRO{\U{ba}}%
%BeginExpansion
{{}^o}%
%EndExpansion
$ with the positive direction of the $\theta$-axis. It should also be noted
that if~$W=-\frac{3}{2}$, the~lines $BB^{\prime}$ and $CC^{\prime}$ coincide,
making an angle of $-33.69%
%TCIMACRO{\U{ba}}%
%BeginExpansion
{{}^o}%
%EndExpansion
$ with the $\theta$-axis.

To continue, let us check if there are regions of the phase diagrams in which
$\rho<0$. In~these regions, the~solutions should not be admitted as physical
solutions, at~least~classically.

We start by replacing $\Psi=\theta\tan\beta$ in (\ref{32}). We thus obtain
\begin{equation}
-\theta^{2}\left(  \frac{W}{2}\tan^{2}\beta-\tan\beta-\frac{1}{3}\right)
=8\pi\rho. \label{67}%
\end{equation}

It is easy to verify, taking into account (\ref{64}), that the invariant rays
lying on the lines $BB^{\prime}$ and $CC^{\prime}$ represent vacuum solutions.
Moreover, we have no region with a negative energy density if $W<-\frac{3}{2}%
$. On~the other hand, when $W>-\frac{3}{2}$, we find regions where $\rho<0$
that are delimited by the invariant rays that lie on the lines $BB^{\prime}$
and $CC^{\prime}$. In~the next section, these regions are represented as
dotted regions in the phase diagrams, which widen as the value of $W$
increases, tending to leave the classically allowed solutions localized in a
narrow region that includes the $\theta$-axis.

\subsection{Phase~Diagrams}

\begin{figure}[h]
\begin{center}
\includegraphics[width=7.5 cm]{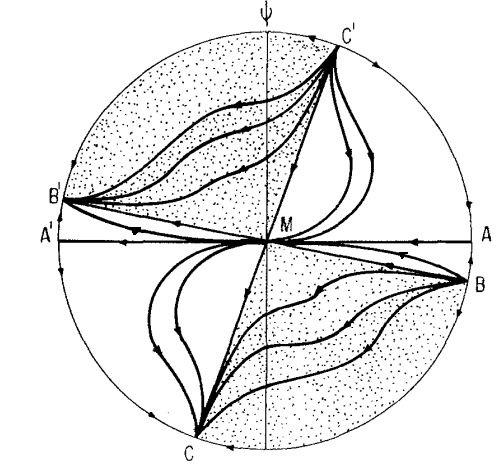}
\end{center}
\unskip
\caption{\centering{$\ \ W>0$ ($\omega>\frac{3}{2} $).}\label{fig3}}
\end{figure}

Now, one can obtain the basic representation of Weyl's cosmological solutions
on the Poincar\'{e} sphere (the phase diagrams). This allows us to make a
qualitative analysis of the solutions at infinity. First, let us make some
comments about the diagrams (\mbox{Figures~\ref{fig1}--\ref{fig3}}), which are
valid for $\lambda\neq1$ and are separated into intervals of $W$ (or $\omega
$)\footnote{The cases $W=-\frac{3}{2},$ $W-\frac{4}{3}$ and $W=0$ were not
analysed because they contain multiple equilibrium points or singularities.}.

Initially, for~$W<-3/2$ (see Figure~\ref{fig1}), the~closed curves appearing
in the diagram represent nonsingular cosmological models, which start in the
infinitely distant past from Minkowski's space-time (the point $M$ $(0,0)$)
and tend to it again in the infinitely distant future; these universes present
an initial phase of contraction, and~then move into an expansive phase.
For~some of these solutions, the~scalar field $\Phi$ is increasing (if
$\Psi>0$), while for the others, it is decreasing, in~which case $\Psi<0$.
On~the other hand, it is possible to have singular solutions with a constant
scalar field ($\Psi=0$): they are represented by the $AM$ curves, which
correspond to solutions that start with a ``big bang'', and~then undergo an
expansive phase, finally tending to Minkowski's space-time, and~the
$MA^{\prime}$ curves, solutions that start from Minkowski's space-time (in the
infinitely distant past, with the cosmic time $t\rightarrow-\infty$),
and~follow a contraction regime until the final~collapse.

In fact, the~curves $AM$ and $MA^{\prime}$ also correspond to solutions of
general relativity, since from (\ref{58}) with $\Psi=0$, it follows that
\begin{equation}
\dot{\theta}=-\frac{\left(  1+\lambda\right)  }{2}\theta^{2}, \label{68}%
\end{equation}
whose solution is%
\begin{equation}
\frac{1}{\theta}=\frac{(1+\lambda)}{2}t+\delta, \label{69}%
\end{equation}
where $\delta$ is an arbitrary constant. Therefore, by~setting $\delta=0$, we
obtain the known scale factor%
\begin{equation}
R(t)=R_{0}t^{2/3(1+\lambda)}. \label{70}%
\end{equation}

In Figure~\ref{fig2}, we consider the interval $-3/2<W<-4/3$. In~this diagram,
there are six invariant rays: $AM$, $MA^{\prime}$, $BM$, $MB^{\prime}$, $CM $,
and $MC^{\prime}$. It is interesting to recall that the dotted regions in the
diagram contain solutions with $\rho<0$, so that the curves restricted to
these regions do not correspond to physical models. Furthermore, solutions
lying on the lines $BB^{\prime}$ and $CC^{\prime}$ are vacuum solutions
($\rho=0$), possessing singularities in their geometries, i.e.,~they are ``big
bang'' models ($BM$ and $CM$) or models that collapse ($MB^{\prime}$and
$MC^{\prime}$), but~with the scalar field varying. In~the region where
$\rho>0$, one finds solutions similar to those in the previous diagram and
also expanding universes with decreasing $\Phi$ ($BM$) and collapsing
universes with increasing $\Phi$ ($MB^{\prime}$).

For $W>-\frac{4}{3}$, it turns out that there are no nonsingular solutions in
the diagrams. In~Figure~\ref{fig3} ($W>0$), in addition to~solutions that
appeared in Figure~\ref{fig2} when $\rho\geq0$, we now observe the existence
of expanding universes with increasing $\Phi$ ($C^{\prime}M$) and collapsing
universes with decreasing $\Phi$ ($MC$). As~mentioned before, if~$W$
increases, the~line $BB^{\prime}$ moves anticlockwise approaching the line
$AA^{\prime}$, while the line $CC^{\prime}$ moves clockwise, also approaching
$AA^{\prime}$; as a consequence, the~``forbidden'' regions (sectors
$MB^{\prime}C^{\prime}$ and $MBC$), where $\rho<0$, become wider. In~the limit
$W\rightarrow\infty$, the~line $AA^{\prime}$ remains in the region where the
energy density $\rho$ is positive, representing the solutions of general
relativity given by (\ref{70}). Actually, for~each value of $W$, the~line
$AA^{\prime}$ contains the solutions (\ref{70}) because $\Phi=const$ (which
implies $\Psi=0$) is a solution to Equation~(\ref{6}).

In most of the diagrams, the equilibrium points do not appear as isolated
points. In these cases, they correspond to multiple equilibrium points,
constituting the invariant rays. In the other cases they appear on the
Poincar\'{e} sphere as points at the infinity, whose nature are indicating in
the table below

\begin{center}%
\begin{tabular}
[c]{|c|c|c|c|}\hline
Intervals & $A,A^{\prime}$ & $B,B^{\prime}$ & $C,C^{\prime}$\\\hline
$W<-3/2$ & saddle points & - & -\\\hline
$-3/2<W<-4/3$ & saddle points & two-tangent nodes & saddle points\\\hline
$W>-4/3$ ($W\neq0$) & saddle points & two-tangent nodes\^{A}%
%TCIMACRO{\U{b4}}%
%BeginExpansion
\'{}%
%EndExpansion
& two-tangent nodes\\\hline
\end{tabular}

\bigskip
\textbf{Table 1}: Behaviour of the equilibrium points on the Poincar\'{e} sphere.

\end{center}

\section{Conclusions}

In this paper, we sought to find cosmological solutions in
the context of the Weyl geometrical scalar--tensor theory. The~vacuum field
equations of this theory are formally identical to those of the Brans--Dicke
theory, so we were able to obtain a Kasner type solution from the
corresponding solution in the Brans--Dicke theory. We also found that, in~the
limit $\omega\rightarrow\infty$, the~Kasner solution of general relativity was
recovered. On~the other hand, we investigated the existence of solutions for
homogeneous and isotropic models sourced by a perfect fluid. In~this case, we
found an analytic solution for stiff matter and~also showed that the
corresponding solution of general relativity could be obtained in the limit
$W\rightarrow\infty$. For~values of the parameter $\lambda\neq1$, no
analytical solution was possible, and~we used dynamical systems theory to
display the phase diagrams of the solutions in intervals of $W$ (or $\omega$).
When $W>0$, we highlighted solutions representing universes with $\rho>0$ and
an increasing geometric scalar field, which started with a ``big bang'' and
expanded to a final phase that tended toward Minkowski's space-time (the
curves $C^{\prime}M$).

An interesting fact regarding the phase diagrams examined here is that there
was no difference between the cosmological models when different values of the
parameter $\lambda$ were considered. In~that sense, it can be seen that
Equation~(\ref{64}), which determines the invariant rays, did not depend on
$\lambda$. Moreover, it should be noted that in the present context, matter
was not a source of the geometric scalar field $\Phi$ in Equation~(\ref{6}).
By~contrast, in~the Brans--Dicke theory, the scalar field equation is%
\begin{equation}
\square\varphi\ =\frac{8\pi T}{2\omega+3}, \label{71}%
\end{equation}
where, as~is well known, $T$ denotes the trace of the energy--momentum tensor.
For the case of a perfect fluid source, $T=T(\lambda)$ and $T=0$ only when
$\lambda=\frac{1}{3}$. As~a consequence, in~the present scalar--tensor theory,
cosmological models differed according to the value of the parameter $\lambda$
\cite{rom89}.

In this work, we did not consider the presence of the cosmological constant,
nor did we take any potential of the scalar field into account. Because~of
this, we did not find any solution describing the acceleration of the
universe. Incidentally, models describing cosmological scenarios in which the
acceleration of the cosmos is driven by a scalar field, quintessence
models~\cite{ste99,toh23} and Chaplygin gas models~\cite{fab17,ben02} among
others~\cite{sen01}, have been investigated with interest. Two lines of
research that we leave for further work are (i) an investigation of the role
the geometric scalar field could play in approaching the problem of dark
matter and (ii) considering scenarios where the cosmological constant is
present with the hope that they can give some light to the problem of
dark~energy. \newline\newline

\textbf{Acknowledgements.} C. Romero thanks CNPq (Brazil) for financial
support. \newline

\end{document}